\documentclass{article}

\usepackage{arxiv}

\usepackage{cite}
\usepackage[utf8]{inputenc} % allow utf-8 input
\usepackage[T1]{fontenc}    % use 8-bit T1 fonts
\usepackage{hyperref}       % hyperlinks
\usepackage{url}            % simple URL typesetting
\usepackage{booktabs}       % professional-quality tables
\usepackage{amsfonts}       % blackboard math symbols
\usepackage{nicefrac}       % compact symbols for 1/2, etc.
\usepackage{microtype}      % microtypography
\usepackage{lipsum}
\usepackage{graphicx}
\usepackage{hyperref}
\usepackage{cleveref}
\usepackage{subcaption}

\graphicspath{ {./images/} }

\title{Cross-Format Retrieval-Augmented Generation in XR with LLMs for Context-Aware Maintenance Assistance}

\author{
1\textsuperscript{st} Akos Nagy
\textit{Dept. of Networks \& Digital Media} \\
\textit{ECE, Kingston University}\\
Kingston upon Thames, UK \\
\texttt{A.Nagy@kingston.ac.uk}
\And
2\textsuperscript{nd} Yannis Spyridis
\textit{Dept. of Computer Science} \\
\textit{ECE, Kingston University}\\
Kingston upon Thames, UK \\
\texttt{Y.Spyridis@kingston.ac.uk}
\And
3\textsuperscript{rd} Vasileios Argyriou
\textit{Dept. of Networks \& Digital Media} \\
\textit{ECE, Kingston University}\\
Kingston upon Thames, UK \\
\texttt{Vasileios.Argyriou@kingston.ac.uk}
}

\begin{document}
\maketitle
\begin{abstract}
This paper presents a detailed evaluation of a Retrieval-Augmented Generation (RAG) system that integrates large language models (LLMs) to enhance information retrieval and instruction generation for maintenance personnel across diverse data formats. We assessed the performance of eight LLMs, emphasizing key metrics such as response speed and accuracy, which were quantified using BLEU and METEOR scores. Our findings reveal that advanced models like GPT-4 and GPT-4o-mini significantly outperform their counterparts, particularly when addressing complex queries requiring multi-format data integration. The results validate the system’s ability to deliver timely and accurate responses, highlighting the potential of RAG frameworks to optimize maintenance operations. Future research will focus on refining retrieval techniques for these models and enhancing response generation, particularly for intricate scenarios, ultimately improving the system's practical applicability in dynamic real-world environments.
\end{abstract}


\section{Introduction}
LLMs \cite{gao2024retrievalaugmentedgenerationlargelanguage} represent a significant leap forward in the field of artificial intelligence (AI) and natural language processing (NLP) \cite{qin2024largelanguagemodelsmeet}, demonstrating unprecedented capabilities in understanding, generating, and manipulating human language. These models, built on deep learning architectures, typically leverage massive datasets and extensive computational resources to capture complex linguistic patterns, allowing them to perform a wide range of tasks, such as text generation, translation, summarization, and question-answering, with remarkable accuracy. LLMs like OpenAI’s GPT series\cite{openai2024gpt4technicalreport} and Meta's Llama models\cite{dubey2024llama3herdmodels}, among others, have set new benchmarks in AI research, pushing the boundaries of what machines can achieve in understanding and interacting with natural language.

As LLMs continue to advance, they are also reshaping the landscape of human-computer interaction. These models have showcased a variety of use-cases and increasingly being integrated into interactive systems where they engage in conversational AI, assistive technologies, and creative collaborations with humans. Their ability to process and generate natural language makes them pivotal in the development of new forms of interaction for intelligent virtual assistants.

In recent years, Retrieval-Augmented Generation (RAG) systems have emerged as a transformative approach within the AI landscape, particularly in enhancing the capabilities of LLMs\cite{articleRamachandran}. By effectively combining retrieval mechanisms with generative models, RAG architectures enable the integration of factual information from various sources \cite{sundar2022multimodalconversationalaisurvey} to produce contextually relevant responses. This capability is particularly advantageous in domains requiring complex decision-making and information synthesis, such as maintenance tasks. In our proposed system, the RAG framework facilitates multi-format data processing, allowing for efficient retrieval from diverse file types, including text documents, CSV files, and databases. This dual approach improves response accuracy by grounding model outputs in real-world information while significantly accelerates the generation process. Thus, our research aims to evaluate and demonstrate the practical implications of RAG systems in optimizing maintenance workflows, and providing timely and comprehensive support to personnel operating in dynamic environments.

The evaluated architecture proposes the following contributions:

\begin{itemize}

    \item Cross-Format Functionality: The system can process and retrieve data from various file types, such as plain text, PDFs, CSVs, and databases. This flexibility allows users to access information from multiple sources without the need for manual conversion, ensuring seamless integration into environments where data is stored in different formats.

    \item Multi-Path Routing: Building on its cross-format capabilities, the system enables simultaneous retrieval from multiple sources in response to a single query. This allows users to access comprehensive information—such as manuals, parts lists, and technical specifications—without needing to search each source individually, improving efficiency and accuracy in maintenance tasks.
    
\end{itemize}

The paper is organized as follows: In Section II, we review related work in the field, highlighting advancements in multimodal interaction systems and their implications for maintenance tasks. Section III details the system design, outlining key components and functionalities of our proposed architecture. In Section IV, we present the evaluation methodology, including the criteria for assessing response speed and accuracy employed in our comparative study of the LLMs. The same section also discusses the results obtained from various scenarios, providing a comprehensive analysis of system performance against the evaluated models. In Section V, we explore the implications of our findings, particularly regarding the integration of human expertise into AI systems and template creation for efficient instruction delivery, and the implication of performance issues regarding certain LLMs. Finally, Section VI concludes the paper, by summarising key insights and outlining future research directions aimed at enhancing the system’s capabilities in real-world applications.

\begin{figure}[t]
\centering
\includegraphics[width=0.5\linewidth]{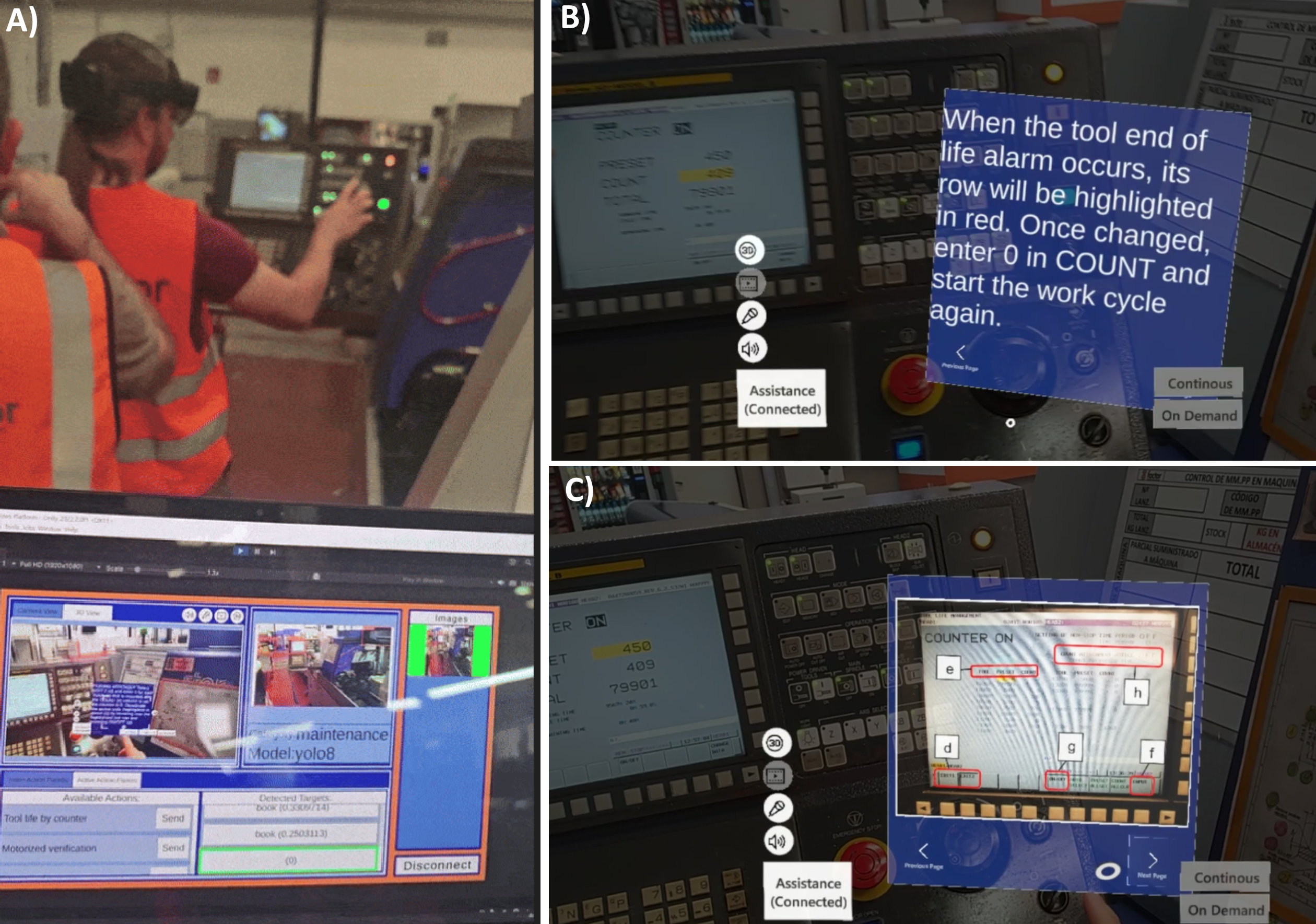}
\caption{Demonstration of Maintenance Application, showing A) Maintenance Personnel using XR application via OHMD while being monitored by Support Personnel using a computer application. The Capabilities of the XR application allow B) textual and C) visual instructions.}
\label{fig:TALON}
\end{figure}

\section{Related Work}

For related works, the paper focuses on contemporary solutions incorporating multi-modal input methodologies to supply information to a LLMs and RAG solutions.

While there are several research studies on enhancing human-computer interaction (HCI) using classical AI-aided solutions\cite{10257282, 10181189}, in recent years, there has been a surge of approaches focusing on advanced multimodal systems and voice-assisted technologies. A notable contribution in this domain is GazePointAR \cite{10.1145/3613904.3642230}, which presents a context-aware voice assistant aimed specifically at wearable augmented reality (AR) devices. This system addresses the common issue of pronoun disambiguation faced by current voice assistants, such as Amazon Alexa and Google Assistant, which often struggle with contextual understanding. By integrating real-time computer vision capabilities to analyze the user's gaze, action and behaviour understanding \cite{BOUR2019289, 6977392}, pointing gestures, 3D representations \cite{4587762, ARGYRIOU20091}, and conversation history, GazePointAR constructs coherent queries that improve the accuracy of responses. This advancement underscores a shift toward utilizing multimodal interactions to create more natural dialogues and efficient HCI experiences.

Continuing the exploration of immersive interfaces, PANDALens \cite{10.1145/3613904.3642320} introduces an AI-assisted narrative documentation system for Optical See-Through Head Mounted Displays (OHMD). Designed to capture life moments in real-time and minimize user distraction, this tool emphasizes proactive engagement through a mixed-initiative interaction model. By analyzing real-time contextual cues, PANDALens provides automatic suggestions and question prompts, thus significantly enhancing the quality of narrative documentation. Nevertheless, the authors acknowledge challenges related to the accurate interpretation of user intent, particularly in dynamic environments, pointing to the unresolved complexities within multimodal AI interactions.

On another front, there is a growing focus on improving explainability and generalizability within autonomous driving systems. The RAG-Driver framework proposed by Yuan et al. \cite{yuan2024ragdrivergeneralisabledrivingexplanations} exemplifies this trend by introducing a Multi-Modal Large Language Model (MLLM) that offers natural language explanations alongside control signals, thus enhancing user understanding of autonomous decision-making processes. This methodology contrasts with previous opaque end-to-end systems, as RAG-Driver facilitates a two-way interaction that bolsters user trust. The results illustrate significant advancements in explainability and performance metrics, laying the groundwork for improved interpretability in autonomous driving contexts.

In the realm of conversational systems, several studies investigate the role of RAG in enhancing user interaction. A suggestion question generator outlined in one paper \cite{10.1145/3589335.3651905} addresses common difficulties users face when formulating queries, proposing a framework to generate relevant suggestion questions based on initial user inputs. This dynamic contextual approach effectively improves conversational flow, revealing a need for systems that facilitate informed user interactions. Complementarily, the work on AI-safe Autocompletion \cite{10.1145/3627673.3679078} seeks to enhance the complexity of autocomplete suggestions in e-commerce search systems by integrating RAG with safety measures and relevance ranking, showcasing practical implications for conversational AI.

Another pivotal addition to the RAG landscape is REAPER (Reasoning-based Retrieval Planner) \cite{10.1145/3627673.3680087}, designed to improve latency and scalability in conversational shopping assistants. By employing a solitary, smaller language model to generate retrieval plans, REAPER enhances the efficiency of multi-step query handling, reducing response generation times dramatically. Through comparative experiments, REAPER's effectiveness is showcased, potentially setting a new benchmark for future applications in complex query contexts.

A broader synthesis of these advancements can be found in the survey on RAG technologies \cite{10.1145/3637528.3671470}, which provides an extensive overview of RAG techniques concerning integration with LLMs. This comprehensive review categorizes existing research into architectures, training strategies, and applications, while also identifying persistent challenges and proposing future research directions aimed at enhancing the reliability and multilingual capabilities of RA-LLMs. Furthermore, RAG-Ex \cite{10.1145/3626772.3657660} introduces a model-agnostic framework that enhances the explainability of RAG systems, addressing the crucial need for transparency in AI outputs to foster user trust. Collectively, these studies reflect a growing emphasis on usability, explainability, and user engagement across various AI-driven environments.

\section{System Design}

\begin{figure}[t]
\centering
\includegraphics[width=1\linewidth]{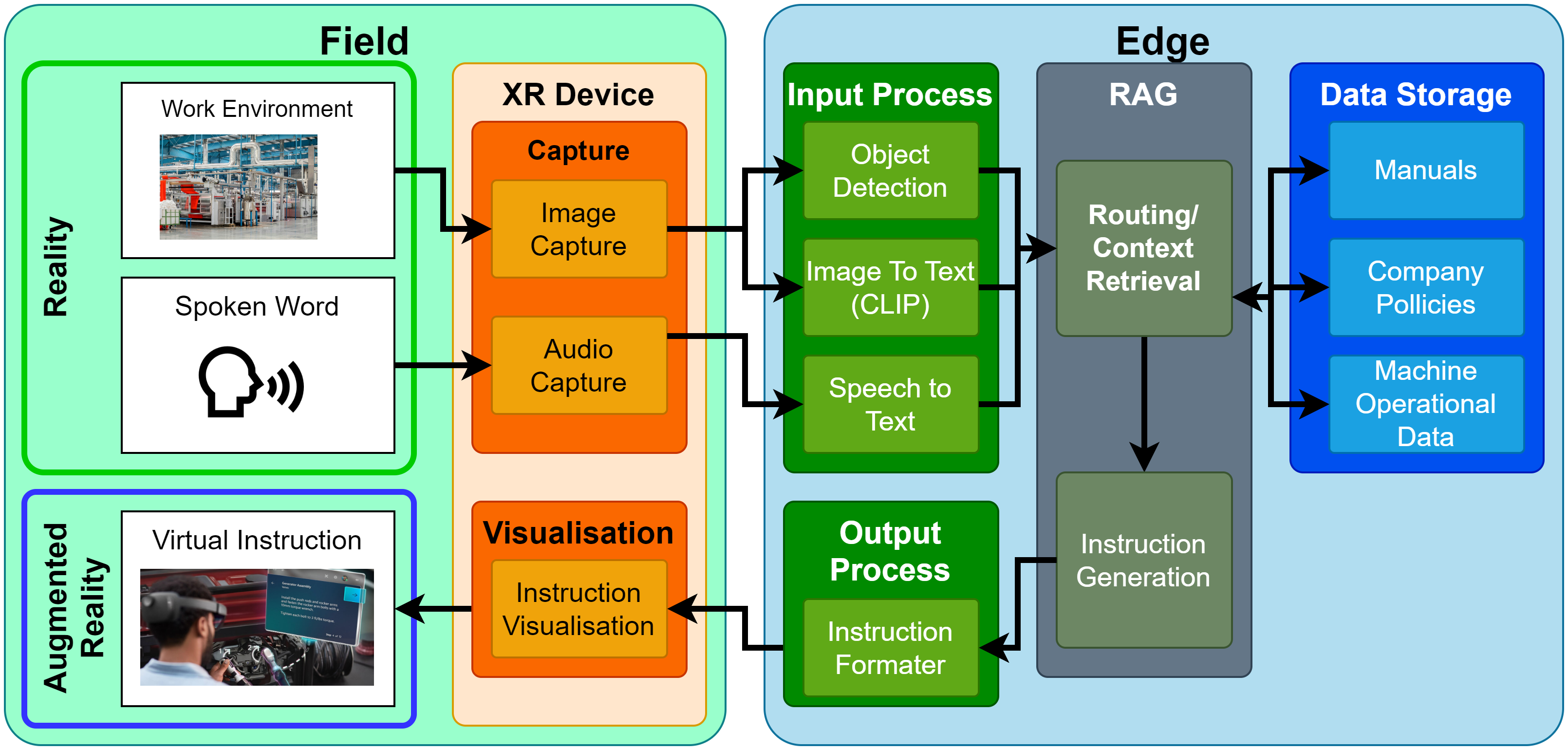}
\caption{Complete Multi-Modal Cross-Format RAG Architecture for Maintenance Procedure Support}
\label{fig:full_architecture}
\end{figure}

\begin{figure}[t]
\centering
\includegraphics[width=1\linewidth]{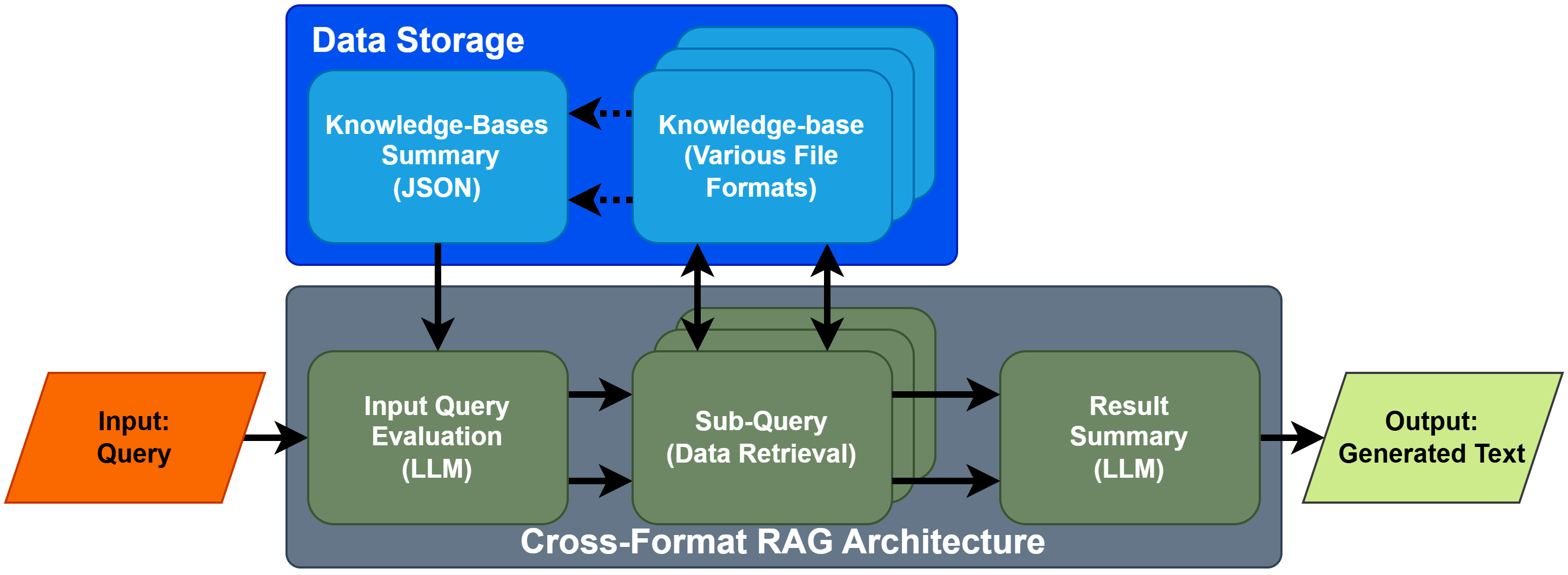}
\caption{RAG Architecture for Cross-Format Data Retrieval}
\label{fig:architecture}
\end{figure}

As the initial Maintenance Support solution \cite{nagy2024userexperienceevaluationar}, as seen on \Cref{fig:TALON}, allows assistance by displaying textual, image, and 3D model information via an OHMD, it relies on the input of external Support Personnel. To decrease the reliance on Support Personnel in routine situations, the proposed architecture, as depicted in \Cref{fig:full_architecture}, is designed to assist Maintenance Personnel in retrieving relevant information pertinent to ongoing maintenance procedures. Users can submit queries or requests as voice recordings, accompanied by images captured in the work environment. To enrich the input data, the system employs AI Vision models that extract pertinent information from the provided images.
This paper evaluates the cross-format capabilities of the RAG architecture, which underpins the data retrieval aspect of the proposed solution. An overview of the RAG architecture is illustrated in \Cref{fig:architecture}.

\textbf{Multi-Path Retrieval}

The system is engineered to handle complex queries that may reference multiple knowledge bases. This feature is crucial for scenarios where multi-part queries relate to various sources or require simultaneous access to multiple information repositories. The architecture utilizes LLMs to analyze input queries and generate subqueries corresponding to the available knowledge bases. This sub-querying mechanism ensures correlation and streamlining across all accessed data, facilitating efficient and relevant data retrieval. The architecture provides a Knowledge Base Summary to the LLM model, presenting a JSON-formatted summary of data contained in the individual knowledge bases. The Multi-Path Retrieval module returns JSON files, a widely understood communication format for LLMs \cite{escarda2024llms, mior2024largelanguagemodelsjson}. The resultant JSON data sets forth all the subqueries and their corresponding knowledge bases, which are subsequently submitted to the Cross-Format Retrieval module for further processing.

\textbf{Cross-Format Knowledge Base Retrieval}

The system encompasses a robust knowledge base that consolidates all relevant operational and company data, thereby ensuring vital resources are readily accessible for maintenance tasks. This knowledge base includes technical manuals, providing detailed documentation for operational procedures, company policies to maintain adherence to organizational guidelines during maintenance, and machine operational data to provide real-time and contextually relevant instructions regarding maintenance and functionality.

The current evaluation focuses on two primary file types:

\begin{itemize}
    \item \textbf{PDF:} These files are parsed using the pypdf \cite{10.1145/3299869.3320212} Python library. Based on the sub-query, the LLM retrieves relevant information from the PDF files.
    
    \item \textbf{CSV:} For Comma Separated Values (CSV) files, the Knowledge Base summary includes the column headers and the first five rows of data from the file, alongside the short summary of the stored data. These details facilitate the LLMs in querying relevant information. The data request process employs the DuckDB \cite{adhikari2024comparativestudypdfparsing} Python library, which allows for SQL-like queries. Depending on the sub-query and the available information about the CSV data, the LLMs generate the requisite SQL code.
\end{itemize}

\section{Evaluation}
The evaluation aimed to assess the performance of our Cross-Format RAG system using different large language models. The study involved 8 LLMs, 4 OpenAI models, and 4 Llama models. It focused on two key criteria: speed of response and accuracy, with the latter evaluated using BLEU and METEOR scores. These metrics provided a quantitative assessment of how closely the system's outputs aligned with the reference solutions, which were defined by humans.

To rigorously assess the performance of our RAG system, we conducted an extensive evaluation across various scenarios and models. Each model was run independently in a controlled setting, with the RAG process executed 50 times per model for each scenario. Importantly, each evaluation was conducted with only one LLM per run, ensuring that the results reflect the capabilities of each individual model without any mixing of models. This structured approach allows for a comprehensive assessment of response speed and accuracy, ultimately providing valuable insights into the effectiveness of our RAG framework as it integrates multiple data formats.

\subsection{Procedure}

The evaluation involved the completion of a set of tasks maintenance-related tasks. The tasks expect an answer or a set of instructions in response to the queries. The information for the correct response was provided in a knowledge-base, consisting of PDF and CSV documents.

\begin{itemize}
    \item \textbf{Scenario A)}  Evaluation of simple queries against a text file-based knowledge source. 
    \item \textbf{Scenario B)} Evaluation of simple queries using data from a CSV file-based knowledge source.
    \item \textbf{Scenario C)} Assessment of complex queries referencing both text and CSV file-based knowledge sources.
\end{itemize}

\subsection{Measures}
\textbf{Success Rate:} The success of a response—whether generated by the RAG system was determined by its ability to fully address the query. A response was considered successful if it provided all the necessary information required to solve the task or issue described in the prompt. This measure ensures that the system's outputs were sufficiently detailed and complete.

The evaluation can end with one of the following results:

\begin{itemize}
    \item Error: An internal issue occurred within the system, preventing any output. 
    \item Incorrect answer: The response did not contain the necessary information.
    \item Partial answer: The response provided incomplete information relevant to the query.
    \item Correct Answer:  The response included all required information as per the query.
    \item Correct answer with additional data: The response included all required information plus extra relevant or related information beyond the query.
    \item Hallucination: The response generated information that was incorrect or misleading, not aligning with the provided query. 
\end{itemize}

\textbf{Speed Measurement:} Speed was a critical factor in the evaluation, and the time taken to produce responses was measured. For the system, the timer started when the prompt was given and ended when the response was generated. By comparing these times, the study aimed to quantify the efficiency of the RAG system in delivering timely solutions in real-world scenarios.

\textbf{Relative Text Length:}  In order to compare the verbosity of the human- and system-generated responses, the study introduced the concept of relative text length. This was expressed as a percentile difference in character count between the two types of responses. By standardising the measurement of text length in this way, the study was able to make meaningful comparisons between responses that varied in length, ensuring that excessively brief or overly detailed outputs could be accounted for in the evaluation of quality and efficiency.

\textbf{Correctness Measurement}: Correctness was evaluated by comparing the responses generated by both the RAG system and human participants against predefined reference answers. Two key metrics were used:

\textbf{BLEU} (Bilingual Evaluation Understudy)\cite{10.3115/1073083.1073135}: BLEU scores were calculated by comparing the system-generated responses to the reference texts, measuring the degree of n-gram overlap. This metric focused on precision, assessing how much of the system's output matched the human-provided reference at different levels of word combinations (n-grams). Higher BLEU scores indicated greater alignment with the reference answer, suggesting that the system produced responses similar to those of human participants in terms of structure and content.

\textbf{METEOR} (Metric for Evaluation of Translation with Explicit ORdering) \cite{banerjee-lavie-2005-meteor}: In addition to BLEU, METEOR was used to evaluate the system’s outputs. METEOR provided a more nuanced measure of correctness by considering factors such as synonymy, stemming, and recall. This metric was particularly valuable in assessing the semantic quality of the responses, as it could identify instances where the system-generated answers used different words but conveyed the same meaning as the reference text. METEOR's inclusion ensured that the system’s performance was evaluated based on its ability to understand and convey the intent behind the prompts.

By combining these metrics, the evaluation provided a comprehensive analysis of how well the RAG system could replicate human-like performance, both in terms of speed and the semantic accuracy of the responses.

\subsection{Results}

\begin{figure*}[ht]
\centering
\includegraphics[width=1\linewidth]{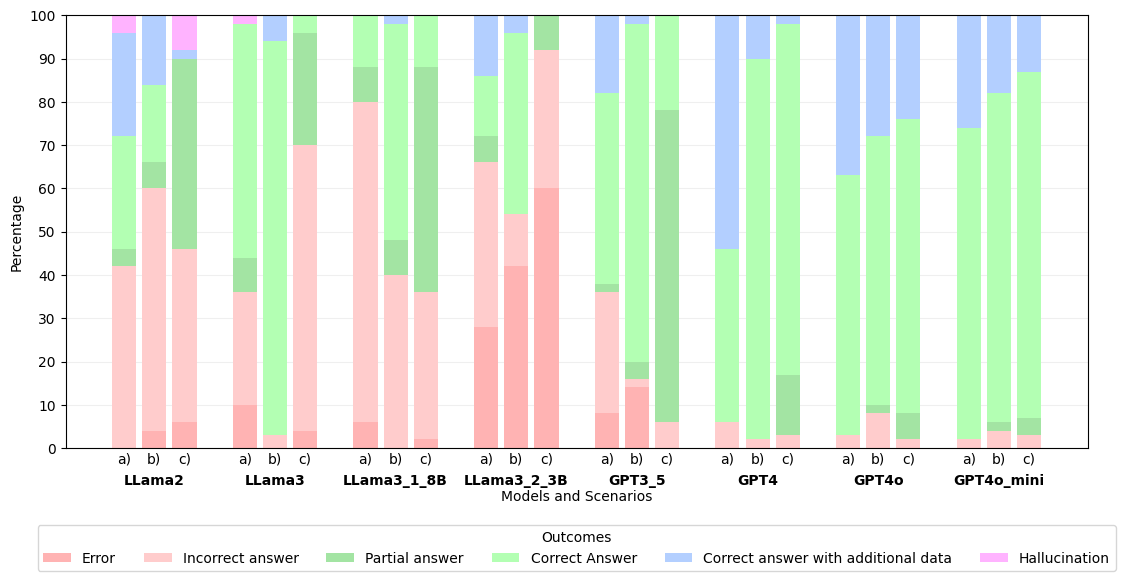}
\caption{Percentage of outcomes of the data generation process per model per scenario}
\label{fig:RagTestStackedDiagram}
\end{figure*}

The evaluation of the RAG system provided comprehensive insights into its performance across different scenarios and models. The results are organized primarily by the three scenarios, each reflecting the complexity of the queries presented to the system and the corresponding responses generated by the various models. Detailed results are shown for Scenario A in \Cref{tab:text_results}, for Scenario B in \Cref{tab:csv_results}, and for Scenario C on \Cref{tab:complex_results}.

\begin{table}[b]
    \centering
    \caption{Summary of Scenario A evaluation results}
    \label{tab:text_results}
    \begin{tabular}{l|rcccc}
        \toprule
            Model & Time & BLEU & METEOR  & Ratio  & Success\\    
        \midrule  
            LLama 2 & 28.42s & 0.14 & 0.38  & 484\% & 34\%\\
            LLama 3 & 6.12s & \textbf{0.49} & \textbf{0.84} & 183\% & 97\%\\
            LLama 3.1:8B & 26.06s & 0.33 & 0.53  & 225\% & 52\%\\
            LLama 3.2:3B & 12.37s & 0.40 & 0.65  & 196\% & 46\%\\
        \midrule  
            GPT-3.5 & \textbf{3.32s} & 0.33 & 0.71 & \textbf{161\%} & 80\%\\
            GPT-4 & 11.58s & 0.18 & 0.58 &  257\% & \textbf{98\%}\\
            GPT-4o & 25.73s & 0.04 & 0.37 &  435\% & 90\%\\
            GPT-4o-mini & 5.22s & 0.17 & 0.60  & 194\% & 94\%\\
        \bottomrule
    \end{tabular}
\end{table}

\begin{table}[t]
    \centering
    \caption{Summary of Scenario B evaluation results}
    \label{tab:csv_results}
    \begin{tabular}{l|rcccc}
        \toprule
            Model & Time & BLEU & METEOR  & Ratio & Success\\      
        \midrule
            LLama 2 & 14.07s & 0.19 & 0.66 & 397\% & 50\% \\
            LLama 3 & 5.52s & 0.09 & 0.51 &  359\% & 54\% \\
            LLama 3.1:8B & 6.22s & 0.08 & 0.47 & 342\% & 12\% \\    
            LLama 3.2:3B & \textbf{2.42s} & 0.31 & 0.58 & 206\% & 28\% \\
        \midrule
            GPT-3.5 & 2.64s & \textbf{0.36} & \textbf{0.74} & \textbf{157\%} & 62\%  \\
            GPT-4 & 4.79s & 0.14 & 0.48 & 191\% & 94\% \\
            GPT-4o & 7.47s & 0.05 & 0.42 & 434\% & 97\% \\
            GPT-4o-mini & 4.41s & 0.07 & 0.41 & 322\% & \textbf{98}\% \\
        \bottomrule
    \end{tabular}
\end{table}

In \textbf{Scenario A}, which evaluated simple queries against a text file-based knowledge source, the results indicated that LLama 3 emerged as the most effective model, achieving a substantial BLEU score of 0.49 and a METEOR score of 0.84, with a success rate of 97%. Notably, it also demonstrated a competitive response time of 6.12 seconds. Conversely, LLama 2 exhibited the weakest performance, recording the lowest BLEU score (0.14) and a success rate of just 34%.

The GPT-4o-mini also showed impressive results, achieving a success rate of 94\% with a response time of 5.22 seconds. In terms of overall speed, GPT-3.5 was the fastest model at 3.32 seconds, albeit with a BLEU score of 0.33 and a success rate of 80\%, indicating a trade-off between speed and the quality of output.

In \textbf{Scenario B}, where queries utilized a CSV file-based knowledge source, the results displayed a shift in model effectiveness. GPT-4o and GPT-4o-mini both produced effective responses, achieving success rates of 97\% and 98\%, respectively. GPT-3.5 outperformed the other large language models with a BLEU score of 0.36 and a METEOR score of 0.74.

On the other hand, LLama 2 had a success rate of only 50\% with a response time of 14.07 seconds, showcasing improved efficiency in the other models while indicating that LLama models generally underperformed with CSV input compared to the more advanced GPT models.

\begin{table}[b]
    \centering
    \caption{Summary of Scenario C evaluation results}
    \label{tab:complex_results}
    \begin{tabular}{l|rcccc}
        \toprule
            Model & Time & BLEU & METEOR  & Ratio & Success\\      
        \midrule
            LLama 2 & 31.28s & 0.06 & 0.30  & 394\% & 2\% \\
            LLama 3 & 10.59s & 0.11 & 0.36 & 211\% & 4\% \\
            LLama 3.1:8B & 12.24s & \textbf{0.19} & \textbf{0.47} & 217\% & 12\%  \\    
            LLama 3.2:3B & \textbf{5.59s} & 0.05 & 0.28 &  286\% & 0\% \\
        \midrule
            GPT-3.5 & 6.21s & 0.10 & 0.25 &  \textbf{93\%} & 22\% \\
            GPT-4 & 19.41s & 0.08 & 0.44 &  209\% & 83\% \\
            GPT-4o & 26.21s & 0.05 & 0.36 &  310\% & 92\% \\
            GPT-4o-mini & 9.90s & 0.10 & \textbf{0.47} & 263\% & \textbf{93}\% \\
        \bottomrule
    \end{tabular}
\end{table}

\textbf{Scenario C} evaluated complex queries requiring referencing both text and CSV data sources. Here, the performance variance among models was even more pronounced. GPT-4o-mini scored the highest with a success rate of 93\% and an average response time of 9.90 seconds. GPT-4 followed closely with a success rate of 83\% and a response time of 19.41 seconds. In contrast, the LLama models struggled significantly, with LLama 2 achieving a success rate of only 2\%.

The reduction in success rates among  LLama models reflects the challenges they face in managing the complexity of queries and drawing from diverse information sources, compared to the more successful GPT models, which consistently produced coherent and relevant results.

An error analysis, as seen in \Cref{fig:RagTestStackedDiagram}, for Scenario C revealed critical insights into the responses generated. The  LLama 3  model had a notable  66\% incorrect answer rate, emphasizing its difficulty in aligning responses with query requirements. Interestingly,  GPT-4o  and  GPT-4o-mini demonstrated low rates of incorrect answers while achieving high rates of generating additional correct information, indicating their capacity to deliver not just accurate but enriched responses. This highlights their potential utility in settings where users require thorough and informative outputs.

Llama2 and Llama3 models had a hallucination rate ranging up to 8\%. This either indicates levels of incorrect data retrieval capabilities or that these models are prone to misalign the generated results when presented with certain types of context. Conversely, GPT-4o and GPT-4o-mini models generate verbose responses, resulting in a generally lower BLEU and METEOR score, despite containing the correct information.

In summary, the results from each experimental scenario demonstrate the strengths of the Retrieval-Augmented Generation system in delivering accurate and timely responses, particularly using advanced models like  GPT-4  and  GPT-4o-mini. As expected, the challenges presented by complex queries amplify the necessity for robust model selection and further refinement of response generation techniques. Overall, the performance exhibited across scenarios indicates a promising foundation for enhancing machine-assisted maintenance processes through intelligent data retrieval and response generation.

\section{Discussion}

 Future work will delve deeper into improving performance across all models while addressing the challenges associated with more complex query handling.

\textbf{Local-Model Concerns}

While Llama models offer the advantage of being deployed locally, safeguarding sensitive operational data by eliminating the need for external data sharing, they are not without performance concerns. Our evaluations revealed marked discrepancies in response accuracy and speed compared to cloud-based models, such as those from OpenAI. These local deployments, while enhancing data security, occasionally struggle to achieve the same benchmark of responsiveness and comprehension showcased by their cloud counterparts. The ability to operate without an internet connection is beneficial in environments with limited connectivity, yet this comes at the cost of access to the extensive and continuously updated knowledge repositories leveraged by cloud-based LLMs. Furthermore, the local resource constraints, including computational power and memory, can impact the efficiency of running sophisticated models like Llama, particularly under high-demand scenarios that require real-time processing of multitiered queries. Consequently, while the local execution ensures tighter data control and compliance with privacy standards, it necessitates a careful balance between maintaining robust model performance and ensuring data security.

\textbf{False-Positives Data Generation}

False positives remain a critical challenge when utilizing state-of-the-art LLMs like those employed in our RAG system. The extensive training datasets used in these models often encompass a vast array of general knowledge, which can lead to instances where the model generates incorrect or misleading outputs that do not accurately represent the specific operational context. These misalignments arise particularly in scenarios involving routine inquiries about common issues, as the model may attempt to draw from its generalized understanding rather than from the curated knowledge base relevant to the task. To mitigate these discrepancies, an adaptive context-based filtering mechanism is essential. This mechanism would ensure that the RAG system consistently prioritizes accurate retrieval from the designated knowledge base over reliance on potentially erroneous training data. By implementing such a system, we could improve the integrity and relevance of the responses generated, thereby enhancing user confidence and satisfaction when seeking instructions or solutions to maintenance-related queries. Establishing guidelines for the selection of contextually appropriate outputs will further aid in aligning model responses with user expectations, substantially reducing false positive occurrences.

\section{Conclusion}
This research provides critical insights into the functionality of a RAG system, that effectively incorporates LLMs to support maintenance tasks. Our evaluations demonstrate that models such as GPT-4 and GPT-4o-mini achieve superior performance in both speed and accuracy, especially in navigating complex, cross-format data environments. These findings highlight the necessity of selecting robust models to fully leverage RAG systems in enhancing operational efficiency across industries increasingly reliant on AI-driven solutions.

Future research endeavours will prioritize refining the system's responsiveness to less common queries while continuing to minimize the reliance on human intervention. By harnessing the strengths of RAG architectures, we aim to facilitate their integration into real-world maintenance processes, creating a seamless interaction between machine intelligence and human operators. Through ongoing improvements, we aspire to enhance the operational capabilities of maintenance personnel, ensuring they receive timely, accurate, and contextually relevant information in their work.

\section*{Acknowledgment}
This work was funded by UK Research and Innovation (UKRI) under the UK government’s Horizon Europe funding guarantee [grant number 10047653] and funded by the European Union [under EC Horizon Europe grant agreement number 101070181 (TALON)].

\bibliography{bib.bib}
\bibliographystyle{ieeetr}

\end{document}